\documentclass[a4paper]{article}
\usepackage[psamsfonts]{amssymb}
\usepackage{amsmath}
\usepackage{pstricks}
\usepackage{cite}
\date{}

\author{Saeed Mirshekari\footnote{smirshekari@ut.ac.ir}\ \ and Amir M. Abbassi\footnote{amabasi@khayam.ut.ac.ir}
\\ {\small Department of Physics, University of Tehran,}
\\ {\small North Karegar Ave., Tehran, Iran.}}
\title{ On Energy Distribution of Two Space-times with Planar and Cylindrical Symmetries }

\begin{document}

\maketitle

\begin{abstract}
Considering encouraging Virbhadra's results about energy
distribution of non-static spherically symmetric metrics in
Kerr-Schild class, it would be interesting to study some
space-times with other symmetries. Using different energy-momentum
complexes, i.e. M\o ller, Einstein, and Tolman, in static
plane-symmetric and cylindrically symmetric solutions of
Einstein-Maxwell equations in 3+1 dimensions, energy (due to
matter and fields including gravity) distribution is studied.
Energy expressions are obtained finite and well-defined.
calculations show interesting coincidences between the results
obtained by Einstein and Tolamn prescriptions. Our results support
the Cooperstock hypothesis about localized energy.

\end{abstract}

\bigskip

\section{Introduction}

Since 1916 when Einstein formulated his general theory of
relativity, one of the greatest unsolved puzzles in the realm of
the theory has been concept of gravitational energy. All that is
said about equivalency of frames and arbitrariness of choosing
coordinates fails here. Although, Einstein had believed that
energy is localized in GR and introduced the first energy-momentum
prescription, but, there is no general agreed definition of energy
in GR, yet.

\medskip

There are three main viewpoints about localization of energy:
localization, non-localization, and quasi-loacalization. Misner et
al. \cite{misner} argued that to look for a local energy-momentum
is looking for the right answer to the wrong question. He showed
that the energy can be localized only in systems which have
spherical symmetry. Cooperstock and Sarracino \cite{cooperstock}
proved that if energy is localizable for spherical systems, then
it can be localized in any system. In 1990, Bondi \cite{bondi}
argued that a non-localizable form of energy is not allowed in GR.
Some physicists propose a new concept in this regard:
\textit{quasi-localization} (for example see \cite{hayward}).
Unlike energy-momentum prescriptions theory, quasi-localization
theory does not restrict one to use particular coordinate system,
but this theory have also its drawbacks \cite{berq}.

\medskip

Besides Einstein \cite{moller} prescription, many other
energy-momentum prescriptions was suggested by different persons
that most well-known ones was given by M\o ller \cite{moller},
Landau-Lifshitz \cite{landau}, Papapetrou \cite{papapetrou},
Bergmann \cite{bergmann}, Tolman \cite{tolman}, and Weinberg
\cite{weinberg}. The idea of energy-momentum prescriptions was
criticized for some reasons. First, using different
energy-momentum prescriptions could lead to different energy
distributions for same space-time. Second, except a few of them
(like M\o ller prescription) for other prescriptions all
calculations must be done in Cartesian coordinate system. Third,
they are non-tensorial (pseudo-tensor) and thus their physical
interpretation seems obscure. However, in 1999, Chang et al.
\cite{chang} showed that every energy-momentum complex can be
associated with distinct boundary term which gives the quasi local
energy-momentum. By this way, he dispels doubts expressed about
the physical meaning of energy-momentum complexes. Finally, we can
define conserved angular momentum quantity only for symmetric
prescriptions \cite{weinberg}, while only a few of them are
symmetric. In fact, anti-symmetric characteristic of Einstein's
prescription was the main motivation for Landau and Lifshitz to
look for an alternative prescription which is symmetric.

\medskip

Gravitation becomes absent in local inertial frames and various
selections of coordinate systems give different results for
gravitational energy. In spite of serious objections to the root
of the subject \cite{misner} there are many attempts to extract
and study the common features of the different proposed
energy-momentum prescriptions. Among these works most cases
discussed were dealt with spherically symmetric metrics,
pioneering by Virbhadra \cite{virbb} who showed that for a general
non-static spherically symmetric metric of Kerr-Schild class
several energy-momentum complexes give the same energy
distribution. Considering a general non-static spherically
symmetric space-time of the Kerr-Schild class, he found a strong
coincidence between different energy-momentum complexes in
Kerr-Schild Cartesian coordinate system. Following this approach,
it seems other types of symmetry including cylindrically and plane
symmetries have been taken of less degree of consideration. Main
contribution of Rosen and Virbhadra \cite{virb_and_Rosen} to study
cylindrical gravitational waves should be addressed.

\medskip

Now in the frame work of just doing comparison between different
energy prescriptions for gravitational fields, here we study the
energy distribution of black plane (Plane-symmetric solution of
Einstein-Maxwell equations) and black string (Cylindrically
symmetric solution of Einstein-Maxwell equations) space-times. The
remainder of the paper is organized as follows. In sections 2, we
introduce the black plane and calculate the energy distribution of
this space-time by M\o ller and Tolman prescriptions. We have
compared these new results with energy distribution obtained by
Einstein energy-momentum prescription. In section 3, we present
the black string and explain about structure of this space-time,
briefly. Using M\o ller, Einstein, and Tolman energy-momentum
prescriptions we have calculated the energy distribution of this
cylindrically symmetric space-time. In the final section we
summarize  the results and present our conclusions.\\\\
\textit{Conventions:} we use geometrized units in which the speed
of light in vacuum $ c $ and the Newtonian gravitational constant
$ G $ are taken to be equal to 1. Through the paper Greek and
Latin indices take values 0..3 and 1..3 respectively.

\bigskip

\section{Black Plane Solution}

The line element in general metric of static plane symmetry is
defined as

\begin{eqnarray}\label{plane metric}
ds^2=-A(r) dt^2+ B(r) dr^2+ C(r) (dx^2+ dy^2)
\end{eqnarray}
Cai and Zhang \cite{Zhang} started with the following action

\begin{eqnarray}\label{action}
S=\frac{1}{16\pi}\int_V d^4x \sqrt{-g}
(R+6\alpha^2-F^{\mu\nu}F_{\mu\nu})-\frac{1}{8\pi}\int_{\partial
V}d^3x\sqrt{-h}K,
\end{eqnarray}
where $R$ is the scalar curvature, $F_{\mu\nu}$ is Maxwell field,
and $\alpha^2=-\frac{\Lambda}{3}>0$ presents the negative
cosmological constant. The quantity $h$ is the induced metric on
$\partial V$, and $K$ its extrinsic curvature. Equations of motion
can be obtained by varying the action (\ref{action}) as follows.

\begin{eqnarray}\label{r3}
G_{\mu\nu}=R_{\mu\nu}-\frac{1}{2}R g_{\mu\nu}&=&8\pi
T^{EM}_{\mu\nu}+3\alpha^2 g_{\mu\nu},\nonumber\\
0&=&\partial _\mu(\sqrt{-g}F^{\mu\nu}),\\
F_{\mu\nu,\rho}+ F_{\nu\rho,\mu}+
F_{\rho\mu,\nu}&=&0,\nonumber\\\nonumber
\end{eqnarray}
where
\begin{eqnarray}
T^{EM}_{\mu\nu}=\frac{1}{4\pi}(F_{\mu\lambda}F_\nu^\lambda-\frac{1}{4}g_{\mu\nu}
F^2),
\end{eqnarray}
is the energy-momentum tensor of the Maxwell field. In the metric
(\ref{plane metric}) solving Eqs. (\ref{r3}), Cai and Zhang found
$A(r)$, $B(r)$, $C(r)$, and $F_{\mu\nu}$. Finally, they give the
line element of this solution, black plane, as \cite{Zhang}

\begin{eqnarray}\label{PAdS}
ds^2=-(\alpha^2r^2-\frac{m}{r}+\frac{q^2}{r^2}) dt^2+
(\alpha^2r^2-\frac{m}{r}+\frac{q^2}{r^2})^{-1}
dr^2+\alpha^2r^2(dx^2+dy^2)
\end{eqnarray}
where $\alpha^2=-\frac{\Lambda}{3}$. $m$ and $q$ are obtained by
using Gauss theorem and Euclidean action method of black membranes
respectively \cite{Cai} respectively as

\begin{eqnarray}
m=-\frac{12 \pi M}{\Lambda}, \:\:\: q=2 \pi Q,
\end{eqnarray}

It should be noted that in (\ref{PAdS}) we have taken $r=\vert
z\vert$ because of the reflection symmetry with respect to the
$z=0$ plane. $Q, M,$ and $\Lambda$ are electric charge density,
ADM mass density, and negative cosmological constant,
respectively. This solution is asymptotically anti-de Sitter not
only in the transverse directions, but also in the membrane
directions \cite{Zhang}. Calculating scaler curvature invariants
it is not difficult to show that space-time (\ref{PAdS}) has a
singularity at $r=0$ plane. This singularity is enclosed by four
event horizons. Causal structure of black plane space-time
Eq.(\ref{PAdS}) is similar to that of Reissner Nordstr\"{o}m black
holes \footnote{Reissner Nordstr\"{o}m anti-de Sitter solution and
its energy distribution is reviewed in the Appendix 1}. In vacuum
background $(M=Q=0)$ line element of (\ref{PAdS}) reduces to

\begin{eqnarray}
ds^2=-(\alpha^2r^2)dt^2+(\alpha^2r^2)^{-1}dr^2+\alpha^2r^2(dx^2+dy^2)
\end{eqnarray}
which is just the plane anti-de Sitter space-time.\\

\bigskip

\subsection{Energy Distribution}

Energy-momentum densities, and energy-momentum 4-vectors in M\o
ller, Einstein, and Tolman prescriptions are presented in Table
\ref{Energy Prescriptions} briefly. Interested reader can refer to
the mentioned references for details. We refer to this table for
doing our calculations.

\medskip

\begin{table}
  \centering
  \caption{Energy-momentum Prescriptions}\label{Energy Prescriptions}
\centering
\begin{tabular}{|c|c|c|}
\hline
&&\\
  Prescription & Energy-momentum Density & Energy-momentum\\

&&\\
\hline
&&\\
  M\o ller \cite{moller} & $M_{i}^{k}=\frac{1}{8\pi}\chi_{i ,l}^{k l}$ &$
P_{i}=\int\int\int M_{i} ^{0}dx^{1}dx^{2}dx^{3}$\\
 & $\chi_{i}^{k l}=\sqrt{-g}(\frac{\partial g_{ip}}{\partial
x^{q}}-\frac{\partial g_{iq}}{\partial x^{p}})g^{kq}g^{lp}$
&$=\frac{1}{8\pi}\int\int\chi_{i}^{0 a}n_{a}ds$\\
&&\\
\hline
&&\\

  Einstein \cite{moller} & $\Theta_{i}^{k}=\frac{1}{16\pi}H_{i ,l}^{k l}$
 & $P_{i}=\int\int\int \Theta_{i} ^{0}dx^{1}dx^{2}dx^{3}$ \\
&$H_{i}^{kl}=-H_{i}^{lk}=\frac{g_{in}}{\sqrt{-g}}[-g(g^{kn}g^{lm}-g^{ln}g^{km})]_{,m}$&$=\frac{1}{16\pi}\int\int H_{i}^{0 a}n_{a}ds$\\
&&\\
\hline
&&\\

 & $T_{i}^{k}=\frac{1}{8\pi}U_{i ,l}^{k l}$
 &$P_{i}=\int\int\int T_{i} ^{0}dx^{1}dx^{2}dx^{3}$\\

Tolman \cite{tolman}&$U_{i}^{k
l}=\sqrt{-g}(-g^{pk}V_{ip}^{l}+\frac{1}{2}g_{i}^{k}g^{pm}V_{pm}^{l})$&$=\frac{1}{8\pi}\int\int U_{i}^{0 a}n_{a}ds$\\
&$V_{jk}^{i}=-\Gamma_{jk}^{i}+\frac{1}{2}g^{i}_{j}\Gamma_{mk}^{m}+\frac{1}{2}g^{i}_{k}\Gamma_{mj}^{m}$&\\
&&\\
\hline
\end{tabular}
\end{table}

Halpern \cite{Halpern} used \textit{Einstein prescription} and
found energy distribution of black planes as:
\begin{eqnarray}\label{Einstein Distribution}
E_{\tiny{E}}=M+\frac{\pi \Lambda Q^2}{3r}-\frac{\Lambda^2 r^3}{36
\pi}
\end{eqnarray}

Considering this result, it would be interesting to study energy
distribution of this space-time by other prescriptions. Here, we
extend his work to M\o ller and Tolman prescriptions.

\medskip

Using \textit{M\o ller prescription} (Table \ref{Energy
Prescriptions}) in space-time (\ref{PAdS}), components of
$\chi_{i}^{jk}$ are obtained as
\begin{eqnarray}\label{chi}
\chi_{t}^{t r}=\frac{\Lambda (2\Lambda r^4-3 m r+6 q^2)}{9r}
\end{eqnarray}
Other components of $\chi_{i}^{jk}$ are equal to zero.
Energy-momentum components are found by surface integral presented
in Table \ref{Energy Prescriptions}. It should be noted that for
black planes, we choose the surface of integration to be a planar
shell with fixed $r$ (and then two fixed values of $z$). So,
surface element, $ds$, is $ds=dx dy$ and $\mu_\sigma$, is unit
radial vector. After integration over planer shell with described
$ds$ and $\mu_\sigma$, we find energy density as
\begin{eqnarray}
E_{\tiny{M}}=P_0=\frac{\chi_t^{t r}}{8\pi}
\end{eqnarray}
Substituting $\chi_t^{t r}$ from Eq.(\ref{chi}), energy density in
black plane space-time (in M\o ller prescription) is obtained
finite and well-define as
\begin{eqnarray}\label{Moller Distribution}
E_{\tiny{M}}=\frac{M}{2}+\frac{\Lambda \pi Q^2}{3
r}+\frac{\Lambda^2 r^3}{36 \pi}
\end{eqnarray}\\

Considering \textit{Tolman prescription} (Teble \ref{Energy
Prescriptions}), we calculate components of super potentials
$U_i^{k l}$ and obtain that only non-zero component is
\begin{eqnarray}\label{Super Tolman}
U_t^{t z}=-\frac{2 \Lambda (\Lambda r^4+3 m r- 3 q^2)}{9 r}
\end{eqnarray}
We use planar shell with fixed $r$ again as the surface of
integration and obtain
\begin{eqnarray}\label{Density Tolman}
E_{\tiny{T}}=P_0=\frac{U_t^{t z}}{8\pi}
\end{eqnarray}
Substituting $U_t^{t z}$ from Eq.(\ref{Super Tolman}) to
Eq.(\ref{Density Tolman}), energy density is given by
\begin{eqnarray}\label{Tolman Distribution}
E_{\tiny{T}}=M+\frac{\Lambda \pi Q^2}{3r}-\frac{\Lambda^2 r^3}{36
\pi}
\end{eqnarray}
which is equal to energy density obtained by Einstein prescription
in the same space-time (Eq.(\ref{Einstein Distribution})).
Radinschi \cite{Radinschi} have obtained same coincidence between
Einstein and Tolman energy-momentum prescriptions for Reissner
Nordstr\"{o}m anti-de Sitter space-time (Appendix). In vacuum
background ($Q=\Lambda=0$) or even when just $\Lambda=0$
Eqs.(\ref{Moller Distribution}),(\ref{Tolman Distribution}) reduce
to $E=\frac{M}{2}$ and $E=M$ respectively.

\medskip

Cai and Zhang \cite{Zhang} found that black plane solution has two
horizons in each side of the plane $z=0$. They defined an extremal
case with only one horizon that occurs if
\begin{eqnarray}\label{Q}
Q=\frac{\sqrt{3}M^{\frac{2}{3}}(-\Lambda)^{\frac{1}{6}}}{2
\pi^{\frac{1}{3}}}
\end{eqnarray}
This only horizon is located at
\begin{eqnarray}\label{r}
r=(9\pi M)^{\frac{1}{3}}(-\Lambda)^{-\frac{2}{3}}
\end{eqnarray}

\medskip

Halpern \cite{Halpern} calculated the energy contained inside this
horizon by Einstein prescription. In this case, from
Eq.(\ref{Tolman Distribution}) and compare it with
Eq.(\ref{Einstein Distribution}) it is obvious that Tolman
prescription also give same result. But, studying M\o ller
prescription in this case with different energy distribution
(Eq.(\ref{Moller Distribution})) would be interesting.
Substituting Eqs.(\ref{Q}),(\ref{r}) in Eq.(\ref{Moller
Distribution}) we obtain total energy contained inside the horizon
as
\begin{eqnarray}
E_{\tiny{ext}}=M(\frac{3}{4}-\frac{3^{\frac{1}{3}}\Lambda^2}{12})
\end{eqnarray}
that for small values of $\Lambda$ leads to
\begin{eqnarray}
E_{\tiny{ext}}=\frac{3}{4}M
\end{eqnarray}

\medskip

However energy distribution in M\o ller prescription is different
from Einstein and Tolman prescriptions in general, but total
energy within the black plane's event horizon for the extremal
case in all three prescriptions is same and equal to
three-quarters of the black plane's mass parameter.\\

\bigskip

\section{Black String Solution}

Static Cylindrically symmetric solution of Einstein-Maxwell
equations is given by following line element
\cite{Zhang},\cite{Lemos}
\begin{eqnarray}\label{black string}
ds^2&=&-(\alpha^2 r^2-\frac{4 M}{\alpha r}+\frac{4 Q^2}{\alpha^2
r^2}) dt^2+(\alpha^2 r^2-\frac{4 M}{\alpha r}+\frac{4
Q^2}{\alpha^2 r^2})^{-1} dr^2\\\nonumber &&+ r^2 d\theta^2+
\alpha^2 r^2 dz^2
\end{eqnarray}
where $\alpha^2=-\frac{\Lambda}{3}$ and $-\infty < t, z<
\infty$,\: $0 \leq r < \infty$,\: $0 \leq \theta \leq 2\pi $. $Q$
and $M$ are the ADM mass and charge per unit length in the $z$
direction, respectively. This space-time is asymptotically anti-de
Sitter in the both of transverse and string directions. Black
string space-time has a singularity at $r=0$ which is enclosed by
two horizons.\\

\bigskip

\subsection{Energy Distribution}

Because of cylindrical symmetry of this space-time we choose a
cylindrical surface surrounding the length $L$ from the string
symmetrically with radius $r$, as the  surface of integration. So,
the infinitesimal surface element is $ds=r d\theta dz$ and normal
unit vector in Cartesian coordinate system ($t, x, y, z$) is
$\mu_\alpha=(0, \frac{x}{r}, \frac{y}{r}, 0)$. If we remain in
polar cylindrical coordinate system ($t, r, \theta, z$) we have
$\mu_\alpha=(0, 1, 0, 0)$. As mentioned in section 1, all
calculations in Einstein and Tolman prescription must be done in
Cartesian coordinate system. In this coordinate system
(considering $\theta=arctan(\frac{y}{x})$, and $r=\sqrt{x^2+y^2}$)
Eq.(\ref{black string}) transform to the following line element.
\begin{eqnarray}\label{Cartesian Black String}
ds^2&=&-\xi(r) dt^2+(\frac{x^2}{r^2 \xi(r)}+\frac{y^2}{r^2}) dx^2+
\frac{2 x y}{r^2} (\frac{1}{\xi(r)}-1) dx dy\\\nonumber
&&+(\frac{y^2}{r^2 \xi(r)}+\frac{x^2}{r^2}) dy^2+ \alpha^2 r^2
dz^2
\end{eqnarray}
where
\begin{eqnarray}
\xi(r)=\alpha^2 r^2-\frac{4 M}{\alpha r}+\frac{4 Q^2}{\alpha^2
r^2}, \:\:\: \alpha^2=-\frac{\Lambda}{3}
\end{eqnarray}

\medskip

Using \textit{M\o ller prescription} (Table \ref{Energy
Prescriptions}) for line element (\ref{Cartesian Black String}) we
obtain following needed components of $\chi_{i}^{j k}$.

\begin{eqnarray}\label{chi string}
\chi_t^{t x}&=&-2 x \frac{\alpha^4 r^{\frac{5}{2}}+2 M \alpha r-4
Q^2 \sqrt{r}}{\alpha r^2}\\\nonumber \chi_t^{t y}&=&-2 y
\frac{\alpha^4 r^{\frac{5}{2}}+2 M \alpha r-4 Q^2 \sqrt{r}}{\alpha
r^2}\\\nonumber
\end{eqnarray}
With Eqs.(\ref{chi string}) and after surface integration over
described cylindrical surface in Cartesian coordinate system we
have
\begin{eqnarray}
E_M&=\frac{1}{8 \pi}&\int_0^L\int_0^{2 \pi} \chi_t^{t \sigma}
\mu_\sigma r d\theta dz\:\:\:\\\nonumber &&=\frac{-\alpha^4 r^4-2
M \alpha r+ 4 Q^2}{2 \alpha r }L\\\nonumber
\end{eqnarray}
It should be noted that similar calculations in polar cylindrical
coordinate system also lead to the same result as expected.

\medskip

Using \textit{Einstein} and \textit{Tolman prescriptions} (Table
\ref{Energy Prescriptions}) for line element (\ref{Cartesian Black
String}), non-zero components of super potentials $H_i^{j k}$, and
$U_i^{j k}$ are obtained as
\begin{eqnarray}\label{ET superpotential}
H_t^{t x}=2 U_t^{t x}&=&\frac{3\alpha^4 r^4-\alpha^2 r^2-12 M
\alpha r+ 12 Q^2}{\alpha r^3}x\\\nonumber H_t^{t x}=2 U_t^{t
y}&=&\frac{3\alpha^4 r^4-\alpha^2 r^2-12 M \alpha r+ 12
Q^2}{\alpha r^3}y\\\nonumber
\end{eqnarray}
With Eq.(\ref{ET superpotential}), and after calculation of the
surface integral over described cylindrical surface in Cartesian
coordinate system, we reach to the same results for both Einstein
and Tolman prescriptions as follows.
\begin{eqnarray}
E_{E, T}&=&\frac{1}{16 \pi}\int_0^L\int_0^{2 \pi} H_t^{t \sigma}
\mu_\sigma r d\theta dz\:\:\:\\\nonumber &&=\frac{1}{8
\pi}\int_0^L\int_0^{2 \pi} U_t^{t \sigma} \mu_\sigma r d\theta
dz\:\:\:\\\nonumber &&=\frac{3 \alpha^4 r^4- \alpha^2 r^2- 12 M
\alpha r+ 12 Q^2}{8 \alpha r}L\\\nonumber
\end{eqnarray}

\bigskip

\section{Conclusion}
Considering black plane and black string space-times, with planar
and cylindrically symmetries respectively, we have studied their
energy distributions by using Einstein, M\o ller, and Tolman
prescriptions. All used prescriptions lead to finite and
well-defined expressions for energy. These reasonable obtained
results also tend to support the Cooperstock hypothesis that the
localized energy is zero for regions where the energy-momentum
tensor vanishes.

\medskip

In the each of black plane and black string space-times, Einstein
and Tolman prescriptions give equal results that can be considered
as an extension of Virbhadra's viewpoint that different
energy-momentum prescriptions may provide some basis to define a
unique quantity. M\o ller prescription give different results due
to pseudo-tensorial nature of energy-momentum complexes.

\medskip

In addition, using M\o ller and Tolman prescriptions, our
calculations on total energy within the black plane's event
horizon for the extremal case leads to interesting results which
are in agreement with previous results in Einstein prescription.
However for black plane space-time energy distribution in M\o ller
prescription is different from Einstein and Tolman prescriptions
in general, but total energy within the black plane's event
horizon for the extremal case in all these prescriptions is same
and equal to three-quarters of the black plane's mass parameter.

\medskip

This work is one of a series of studies by the authors on energy-momentum prescriptions in general relativity~\cite{MA1, MA2, MA3}. 

\bigskip

\begin{center}
\textbf{\begin{Large}Appendix\end{Large}} \linebreak

\begin{large}\textbf{RN-AdS Solution and its Energy
Distribution}\end{large}
\end{center}

Since casual structure of black plane space-time is similar to
that of Reissner-Nordstr\"{o}m anti-de Sitter (RN-AdS) space-time,
we are interested to study this space-time more. General form of
this metric is given by this line element \cite{Mann}
\begin{eqnarray}
ds^2=-N(r) dt^2+ \frac{dr^2}{N(r)}+ r d\Omega_b^2\nonumber
\end{eqnarray}
where
\begin{eqnarray}
N(r)=-s\frac{r^2}{\alpha^2}+ b -\frac{2
m}{r}+\frac{q^2}{r^2}\nonumber
\end{eqnarray}
with $\alpha^2=-\frac{3}{\vert \Lambda \vert}$, $s=\frac{\vert
\Lambda \vert}{\Lambda}$ is the sign of $\Lambda$, and

\begin{eqnarray}
d\Omega_b=
\left\{%
\begin{array}{ll}
    d\theta^2+ \sin^2{\theta} d\phi^2, & \hbox{$b=1$, $s=\pm 1$;} \\\nonumber
    d\theta^2+ d\phi^2, & \hbox{$b=0$, $s=-1$;} \\\nonumber
    d\theta^2+ \sinh^2{\theta} d\phi^2, & \hbox{$b=-1$, $s=-1$.} \\\nonumber
\end{array}%
\right.
\end{eqnarray}
For $b=1$, above metric describes the RN-AdS black holes. The
event horizon of the black hole has the 2-sphere topology $S^2$,
and the topology of the space-time is $R^2 \times S^2$
\cite{Radinschi}. Energy distribution of this metric in the case
of $b=1$ in Einstein and Tolman prescriptions are same and given
by \cite{Radinschi}

\begin{eqnarray}
E_{\tiny{E, T}}(r)=M-\frac{Q^2}{2r}+\frac{1}{6}\Lambda
r^3\nonumber
\end{eqnarray}
but in M\o ller prescription we have \cite{Salti}
\begin{eqnarray}
E_{\tiny{M}}(r)=M-\frac{Q^2}{r}-3\frac{r^3}{\Lambda}\nonumber
\end{eqnarray}

\end{document}